# Unsymmetrical and symmetrical one-range addition theorems for Slater type orbitals and Coulomb-Yukawa like correlated interaction potentials of integer and noninteger indices

I.I. Guseinov

*Depertmant of Physics, Faculty of Arts and Sciences, Onsekiz Mart University, Çanakkale,Turkey*

**Abstract**

Using one-center expansion relations for the Slater type orbitals (STOs) of noninteger principal quantum numbers in terms of integer n STOs derived in this study with the help of $\Psi^\alpha$- exponential type orbitals ($\Psi^\alpha$-ETOs , $\alpha = 1, 0, -1, -2,...$) the general formulas are established for the unsymmetrical and symmetrical one-range addition theorems of STOs and Coulomb-Yukawa like correlated interaction potentials (CIPs) with integer and noninteger indices. The final results are especially useful for computations of arbitrary multicenter multielectron integrals over STOs that arise in the Hartree-Fock-Roothaan (HFR) approximation and also in the correlated methods which play a significant role in theory and application to quantum mechanics of atoms, molecules, and solids.

## I. Introduction

The multicenter multielectron integrals are of fundamental importance in the determination of multielectron properties for atoms and molecules when the Hylleraas approach in Hartree-Fock theory is employed [1]. However, the difficulties in calculation of these integrals have restricted the application of Hylleraas approximation in quantum chemistry. In the literature there is renewed interest in developing efficient methods for the calculation of multicenter multielectron integrals over STOs (see, e.g., [2-5] and references quoted therein). Older work are reviewed in [6,7]. It should be noted that the preexisting formulas for the evaluation of multicenter multielectron integrals do not generally apply to arbitrary correlated interaction potentials. One of the most promising correlated methods is based upon the use of one-range addition theorems for correlated interaction potentials approach suggested in our previous papers using complete orthonormal sets of $\psi^\alpha$-ETOs [8]. In Ref. [9], by the use of complete orthonormal sets of $\Psi^\alpha$-ETOs we presented a particular method for obtaining the unsymmetrical one-range addition theorems for STOs and correlated interaction potentials of integer indices which has been utilized for the evaluation of multicenter multielectron integrals of Coulomb-Yukawa like CIPs in terms of two-center overlap integrals with the same screening parameters. The aim of this work is to establish the series expansion formulas for the unsymmetrical and symmetrical one-range addition theorems for STOs and Coulomb-Yukawa like CIPs with arbitrary integer and noninteger indices which can be useful in the evaluation of multicenter multielectron integrals that arise in the solution of multielectron atomic and molecular problems when a correlated interaction potentials approximation in HFR theory is employed.

## 2. One-center expansion of STOs and CIPs of noninteger indices

The noninteger $n^*$ STOs and noninteger $\mu^*$ CIPs are defined as

$$\chi_{n^*lm}(\zeta^*,\vec{r}) = (2\zeta^*)^{n^*+1/2}[\Gamma(2n^*+1)]^{-1/2} r^{n^*-1} e^{-\zeta^* r} S_{lm}(\theta,\varphi) \tag{1}$$

$$f_{\mu^*\nu\sigma}(\xi^*,\vec{r}) = \left(\frac{4\pi}{2\nu+1}\right)^{1/2} r^{\mu^*-1} e^{-\xi^* r} S_{\nu\sigma}(\theta,\varphi), \tag{2}$$

where $S_{lm}$ and $\Gamma(x)$ are complex $(S_{lm} \equiv Y_{lm})$ or real spherical harmonics and the gamma function, respectively. The integer n STOs and integer $\mu$ CIPs are obtained from Eqs. (1) and (2) for $n^* = n$ and $\mu^* = \mu$, where n and $\mu$ are the integers,

$$\chi_{nlm}(\zeta,\vec{r}) = (2\zeta)^{n+1/2}[(2n)!]^{-1/2} r^{n-1} e^{-\zeta r} S_{lm}(\theta,\varphi) \tag{3}$$

$$f_{\mu\nu\sigma}(\xi,\vec{r}) = \left(\frac{4\pi}{2\nu+1}\right)^{1/2} r^{\mu-1} e^{-\xi r} S_{\nu\sigma}(\theta,\varphi). \tag{4}$$

In order to establish the one-range addition theorems for STOs and CIPs of noninteger indices we shall utilize their one-center expansion relations. For this purpose we use the method set out in [8] and properties of $\Psi^\alpha$-ETOs. Then, we find finally for the one-center expansion in terms of integer n STOs and integer $\mu$ CIPs the following relations:

for noninteger $n^*$ STOs

$$\chi_{n^*lm}(\zeta^*,\vec{r}) = \lim_{M\to\infty} \sum_{n=l+1}^{M} P^{\alpha M}_{n^*l,nl}(t^*) \chi_{nlm}(\zeta,\vec{r}) \tag{5}$$

$$P^{\alpha M}_{n^*l,nl}(t^*) = \sum_{n'=l+1}^{M} \Omega^{\alpha l}_{nn'}(M) \frac{\Gamma(n^*+n'-\alpha+1)}{[\Gamma(2n^*+1)\Gamma(2n'-2\alpha+1)]^{1/2}} (1+t^*)^{n^*+1/2}(1-t^*)^{n'-\alpha+1/2}, \tag{6}$$

for noninteger $\mu^*$ CIPs

$$f_{\mu^*v\sigma}(\xi^*,\vec{r}) = \lim_{M\to\infty}\sum_{\mu=v+1}^{M} Q^{\alpha M}_{\mu^*v,\mu v}(\tau^*) f_{\mu v\sigma}(\xi,\vec{r}) \tag{7}$$

$$Q^{\alpha M}_{\mu^*v,\mu v}(\tau^*) = \frac{1}{(\xi^*+\xi)^{\mu^*+1/2}} \sum_{\mu'=v+1}^{M} \Omega^{\alpha v}_{\mu\mu'}(M) \frac{\Gamma(\mu^*+\mu'-\alpha+1)}{[\Gamma(2\mu'-2\alpha+1)]^{1/2}} (1-\tau^*)^{\mu'-\alpha+1/2}. \tag{8}$$

Here $-\infty < \alpha \leq 1$, $\zeta^* > 0$, $\zeta > 0$, $\xi^* \geq 0$, $\xi > 0$, $t^* = \frac{\zeta^*-\zeta}{\zeta^*+\zeta}$, $\tau^* = \frac{\xi^*-\xi}{\xi^*+\xi}$ and

$$\Omega^{\alpha l}_{nk}(N) = \left[\frac{[2(k-\alpha)]!}{(2k)!}\right]^{1/2} \sum_{n'=\max(n,k)}^{N} (2n')^{\alpha} \omega^{\alpha l}_{n'n} \omega^{\alpha l}_{n'k}, \tag{9}$$

$$\omega^{\alpha l}_{nn'} = (-1)^{n'-l-1} \left[\frac{(n'+l+1)!}{(2n)^{\alpha}(n'+l+1-\alpha)!} F_{n'+l+1-\alpha}(n+l+1-\alpha) F_{n'-l-1}(n-l-1) F_{n'-l-1}(2n')\right]^{1/2}, \tag{10}$$

where $F_k(n) = n!/k!(n-k)!$ is the binomial coefficient.

It is easy to show that in the case of integer values of $n^*(n^*=n)$ and $\mu^*(\mu^*=\mu)$ the expansion coefficients occurring in Eqs. (5) and (7) for $t^*=\tau^*=0$ are reduced to the Kronecker symbols, i.e.,

$$P^{\alpha M}_{n^*l,nl}(0) = \delta_{n^*n}\delta_{Mn} \tag{11}$$

$$Q^{\alpha M}_{\mu^*v,\mu v}(0) = \delta_{\mu^*\mu}\delta_{M\mu}. \tag{12}$$

### 3. Expressions for one-range addition theorems of STOs and CIPs

By the use of method set out in previous paper [8] and Eqs. (5) and (7) for the one-center expansion formulas we obtain the following relations:

unsymmetrical one-range addition theorems:

$$\begin{aligned}\chi_{n^*lm}(\zeta^*,\vec{r}_{a1}) &= \lim_{M\to\infty}\sum_{n=l+1}^{M} P^{\alpha M}_{n^*l,nl}(t^*)\chi_{nlm}(\zeta,\vec{r}_{a1}) \\ &= \lim_{\substack{M\to\infty\\N\to\infty}}\sum_{n=l+1}^{M} P^{\alpha M}_{n^*l,nl}(t^*)\sum_{n'=1}^{N}\sum_{l'=0}^{n'-1}\sum_{m'=-l'}^{l'} V^{\alpha N}_{nlm,n'l'm'}(\zeta,\zeta';\vec{R}_{ab})\chi_{n'l'm'}(\zeta',\vec{r}_{b1})\end{aligned} \tag{13}$$

$$f_{\mu^*v\sigma}(\xi^*,\vec{r}_{a1}) = \lim_{M\to\infty}\sum_{\mu=v+1}^{M} Q^{\alpha M}_{\mu^*v,\mu v} f_{\mu v\sigma}(\xi,\vec{r}_{a1})$$

$$= \lim_{\substack{M\to\infty \\ N\to\infty}}\sum_{\mu=v+1}^{M} Q^{\alpha M}_{\mu^*v,\mu v}(\tau^*)\sqrt{4\pi}\sum_{\mu'=1}^{N}\sum_{v'=0}^{\mu'-1}\sum_{\sigma'=-v'}^{v'} W^{\alpha N}_{\mu v\sigma,\mu'v'\sigma'}(\xi,\xi';\vec{R}_{ab})\chi_{\mu'v'\sigma'}(\xi',\vec{r}_{b1}),\qquad(14)$$

where

$$V^{\alpha N}_{nlm,n'l'm'}(\zeta,\zeta';\vec{R}_{ab}) = \sum_{n''=l'+1}^{N} \Omega^{\alpha l'}_{n'n''}(N) S_{nlm,n''-\alpha l'm'}(\zeta,\zeta';\vec{R}_{ab}),\qquad(15)$$

$$W^{\alpha N}_{\mu v\sigma,\mu'v'\sigma'}(\xi,\xi';\vec{R}_{ab}) = \sum_{\mu''=\mu'+1}^{N} \Omega^{\alpha v'}_{\mu'\mu''}(N)\Lambda_{\mu v\sigma,\mu''-\alpha v'\sigma'}(\xi,\xi';\vec{R}_{ab}).\qquad(16)$$

Here, the quantities $S_{nlm,n''-\alpha l'm'}(\zeta,\zeta';\vec{R}_{ab})$ and $\Lambda_{\mu v\sigma,\mu''-\alpha v'\sigma'}(\xi,\xi';\vec{R}_{ab})$ are the overlap integrlas of integer indices defined as

$$S_{nlm,n''-\alpha l'm'}(\zeta,\zeta';\vec{R}_{ab}) = \int \chi^*_{nlm}(\zeta,\vec{r}_{a1})\chi_{n''-\alpha l'm'}(\zeta',\vec{r}_{b1})dv_1 \qquad(17)$$

$$\Lambda_{\mu v\sigma,\mu''-\alpha v'\sigma'}(\xi,\xi';\vec{R}_{ab}) = \frac{1}{\sqrt{4\pi}}\int f^*_{\mu v\sigma}(\xi,\vec{r}_{a1})\chi_{\mu''-\alpha v'\sigma'}(\xi',\vec{r}_{b1})dv_1.\qquad(18)$$

Symmetrical one-range addition theorems:

$$\chi_{n^*lm}(\zeta^*,\vec{r}_{a1}) = \lim_{M\to\infty}\sum_{n=l+1}^{M} P^{\alpha M}_{n^*l,nl}(t^*)\chi_{nlm}(\zeta,\vec{r}_{a1})$$

$$= \lim_{\substack{M\to\infty \\ N\to\infty \\ N'\to\infty}}\sum_{n=l+1}^{M} P^{\alpha M}_{n^*l,nl}(t^*)\frac{1}{\zeta'^{3/2}}\sum_{n'=1}^{N}\sum_{l'=0}^{n'-1}\sum_{m'=-l'}^{l'}\{\sum_{u=1}^{N+N'-\alpha+1}\sum_{v=0}^{u-1}\sum_{s=-v}^{v} D^{\alpha uvs}_{nlm,n'l'm'}(N,N';\zeta,\zeta')\chi^*_{uvs}(\zeta',\vec{R}_{ab})\}\chi_{n'l'm'}(\zeta',\vec{r}_{b1}) \qquad(19)$$

$$f_{\mu^*v\sigma}(\xi^*,\vec{r}_{21}) = \lim_{M\to\infty}\sum_{\mu=v+1}^{M} Q^{\alpha M}_{\mu^*v,\mu v}(\tau^*)f_{\mu v\sigma}(\xi,\vec{r}_{21})$$

$$= \lim_{\substack{M\to\infty \\ N\to\infty \\ N'\to\infty}}\sum_{\mu=v+1}^{M} Q^{\alpha M}_{\mu^*v,\mu v}(\tau^*)\frac{2^{3/2}}{(2\xi')^{\mu+2}}\sum_{\mu'=1}^{N}\sum_{v'=0}^{\mu'-1}\sum_{\sigma'=-v'}^{v'}\{\sum_{u=1}^{N+N'-\alpha+1}\sum_{v=0}^{u-1}\sum_{s=-v}^{v} B^{\alpha uvs}_{\mu v\sigma,\mu'v'\sigma'}(N,N';\xi,\xi')\chi^*_{uvs}(\xi',\vec{r}_2)\}\chi_{\mu'v'\sigma'}(\xi',\vec{r}_1). \qquad(20)$$

See Ref. [10] for the exact definition of coefficients $D^{\alpha uvs}_{nlm,n'l'm'}(N,N';\zeta,\zeta')$ and $B^{\alpha uvs}_{\mu v\sigma,\mu'v'\sigma'}(N,N';\xi,\xi')$. It should be noted that, in the case of Coulomb like CIPs (for $\xi=0$) the coefficients $B^{\alpha uvs}_{\mu v\sigma,\mu'v'\sigma'}(N,N';\xi,\xi')$ do not depend on the parameter $\xi'$, i.e., $B^{\alpha uvs}_{\mu v\sigma,\mu'v'\sigma'}(N,N';0,\xi') \equiv B^{\alpha uvs}_{\mu v\sigma,\mu'v'\sigma'}(N,N')$.

As can be seen from the formulas obtained in this work, all the one-range addition theorems are expressed through the integer n STOs.

We notice that the unsymmetrical and symmetrical one-range addition theorems of Coulomb and Yukawa potentials are obtained from Eqs. (14) and (20) for ($\mu^* \neq \mu : \nu = \sigma = 0, \xi^* = 0, \xi \neq 0$ or $\mu^* = \mu : \mu = \nu = \sigma = 0, \xi^* = \xi = 0$) and ($\mu^* \neq \mu : \nu = \sigma = 0, \xi^* \neq 0, \xi \neq 0$ or $\mu^* = \mu : \mu = \nu = \sigma = 0, \xi^* = \xi \neq 0$), respectively.

Thus, we have derived a large number of different ($\alpha = 1, 0, -1, -2, ...$) sets of one-range addition theorems for STOs and Coulomb-Yukawa like CIPs of integer and noninteger indices which can be chosen properly according to the nature of the problems under consideration. This is rather important because the choice of the addition theorem set will determine the rate of convergence of the resulting series expansions. Using addition theorems obtained in this study it is easy to show that the arbitrary multicenter multielectron integral with integer and noninteger indices of STOs and interaction potentials that arises in the solution of atomic and molecular problems occurring in HFR and explicitly correlated theories can be expressed in terms of two- and three- center overlap integrals of three integer n STOs. Therefore, the elaboration of the HFR theory with STOs and Coulomb-Yukawa like CIPs of integer and noninteger indices necessitates progress in the development of methods to calculate two- and three- center overlap integrals of three integer n STOs (see Refs. [11- 12]).